\documentclass[aps,prb,twocolumn,letterpaper]{revtex4}

\usepackage{amsmath,amssymb,amsfonts,amsthm}
\usepackage{graphicx}

\begin{document}

\title{Magnetic coupling in CoCr$_2$O$_4$ and MnCr$_2$O$_4$: an LSDA+$U$
  study}

\date{\today}

\author{Claude Ederer}
\affiliation{Department of Physics, Columbia University, 538 West 120th
  Street, New York, NY 10027, U.S.A.}  \email{ederer@phys.columbia.edu}
\author{Matej Komelj}
\affiliation{Jo\v zef Stefan Institute, Jamova 39, SI-1000 Ljubljana,
  Slovenia}
 
\begin{abstract}
  We present a first principles LSDA+$U$ study of the magnetic
  coupling constants in the spinel magnets CoCr$_2$O$_4$ and
  MnCr$_2$O$_4$. Our calculated coupling constants highlight the
  possible importance of $AA$ interactions in spinel systems with
  magnetic ions on both $A$ and $B$ sites.  Furthermore, we show that
  a careful analysis of the dependence of the magnetic coupling
  constants on the LSDA+$U$ parameters provides valuable insights in
  the underlying coupling mechanisms, and allows to obtain a
  quantitative estimate of the magnetic coupling constants. We discuss
  in detail the capabilities and possible pitfalls of the LSDA+$U$
  method in determining magnetic coupling constants in complex
  transition metal oxides.
\end{abstract}


\maketitle

\section{Introduction}

Geometrically frustrated spin systems exhibit a low-temperature
behavior that is fundamentally different from conventional
(non-frustrated) spin systems.\cite{Ramirez:2001, Ramirez:2003} The
incompatibility between local interactions and global symmetry in
geometrically frustrated magnets leads to a macroscopic degeneracy
that prevents these systems from ordering. In some cases this
degeneracy is lifted by further neighbor interactions or by a
symmetry-breaking lattice distortion, resulting in ordered spin
structures at temperatures that are significantly lower than what
would be expected simply from the strength of the nearest neighbor
interaction. Since usually several different ordered configurations
with comparable energy exist in these systems, a very rich low
temperature phase diagram can be observed.

Recently, it has been found in various magnetic spinel systems
(general chemical formula: $AB_2X_4$) that the geometrical frustration
among the $B$ sites in the spinel structure can give rise to
pronounced effects due to \emph{spin-lattice coupling}. In
ZnCr$_2$O$_4$ and CdCr$_2$O$_4$ the macroscopic degeneracy is lifted
by a tetragonal lattice distortion, resulting in complicated
non-collinear spin ordering.\cite{Lee_et_al:2000, Chung_et_al:2005} In
addition, a pronounced splitting of certain phonon modes due to strong
\emph{spin-phonon coupling} has been found in
ZnCr$_2$O$_4$.\cite{Sushkov_et_al:2005,Fennie/Rabe:2006} Non-collinear
spiral magnetic ordering at low temperatures has also been found in
CoCr$_2$O$_4$ and MnCr$_2$O$_4$,\cite{Hastings/Corliss:1962,
Menyuk/Dwight/Wold:1964} where the presence of a second magnetic
cation on the spinel $A$ site lifts the macroscopic degeneracy. Such
non-collinear spiral magnetic order can break spatial inversion
symmetry and lead to the appearance of a small electric polarization
and pronounced \emph{magneto-electric
coupling}.\cite{Kimura_et_al_Nature:2003, Lawes_et_al:2005} Indeed,
dielectric anomalies at the magnetic transition temperatures have been
found in polycrystalline CoCr$_2$O$_4$, \cite{Lawes_et_al:2006} and
recently a small electric polarization has been detected in single
crystals of the same material.\cite{Yamasaki_et_al:2006} Magnetic
spinels therefore constitute a particularly interesting class of
frustrated spin systems exhibiting various forms of coupling between
their magnetic and structural properties.  Furthermore, both $A$ and
$B$ sites in the spinel structure can be occupied by various magnetic
ions and simultaneously the $X$ anion can be varied between O, S, or
Se. This compositional flexibility opens up the possibility to
chemically tune the properties of these systems.

To understand the underlying mechanisms of the various forms of
magneto-structural coupling, it is important to first understand the
complex magnetic structures found in these systems. Such complex
magnetic structures can be studied using model Hamiltonians for
interacting spin systems, which can be treated either classically or
fully quantum mechanically. For the cubic spinel systems, a theory of
the ground state spin configuration has been presented by Lyons,
Kaplan, Dwight, and Menyuk (LKDM)\cite{Lyons_et_al:1962} about 45
years ago. Using a model of classical Heisenberg spins and considering
only $BB$ and $AB$ nearest neighbor interactions, LKDM could show that
in this case the ground state magnetic structure is determined by the
parameter
\begin{equation}
u = \frac{4 \tilde{J}_{BB} S_B}{3 \tilde{J}_{AB} S_A} \quad ,
\end{equation}
which represents the relative strength between the two different nearest
neighbor interactions $\tilde{J}_{BB}$ and $\tilde{J}_{AB}$.\cite{footnote1}
For $u \leq u_0=8/9$ the collinear N{\'e}el configuration, i.e.  all $A$-site
spins parallel to each other and anti-parallel to the $B$-site spins, is the
stable ground state.  For $u > u_0$ it was shown that a ferrimagnetic spiral
configuration has the lowest energy out of a large set of possible spin
configurations and that it is locally stable for $u_0 < u < u''\approx 1.298$.
For $u > u''$ this ferrimagnetic spiral configuration is unstable.  Therefore,
it was suggested that the ferrimagnetic spiral is very likely the ground state
for $u_0 < u < u''$, but can definitely not be the ground state for $u >
u''$.\cite{Lyons_et_al:1962}

On the other hand it has been found that neutron scattering data for both
CoCr$_2$O$_4$ and MnCr$_2$O$_4$ are well described by the ferrimagnetic spiral
configuration suggested by LKDM, although a fit of the experimental data to
the theoretical spin structure leads to values of $u \approx 2.0$ for
CoCr$_2$O$_4$,\cite{Menyuk/Dwight/Wold:1964} and $u \approx 1.6$ for
MnCr$_2$O$_4$,\cite{Hastings/Corliss:1962} which according to the LKDM theory
correspond to the locally unstable regime. Surprisingly, the overall agreement
of the fit is better for CoCr$_2$O$_4$ than for MnCr$_2$O$_4$, even though the
value of $u$ for CoCr$_2$O$_4$ is further within the unstable region than in
the case of MnCr$_2$O$_4$.  From this it has been concluded that: i) the
ferrimagnetic spiral is a good approximation of the true ground state
structure even for $u > u''$, ii) that the importance of effects not included
in the theory of LKDM is probably more significant in MnCr$_2$O$_4$ than in
CoCr$_2$O$_4$, and iii) that the ferrimagnetic spiral is indeed very likely to
be the true ground state for systems with $u_0 < u <
u''$.\cite{Hastings/Corliss:1962,Menyuk/Dwight/Wold:1964,Lyons_et_al:1962}

Recently, Tomiyasu {\it et al.} fitted their neutron scattering data
for CoCr$_2$O$_4$ and MnCr$_2$O$_4$ using a ferrimagnetic spiral
structure similar to the one proposed by LKDM but with the cone angles
of the individual magnetic sublattices not restricted to the LKDM
theory.\cite{Tomiyasu/Fukunaga/Suzuki:2004} As originally suggested by
LKDM, they interpreted their results as indicative of a collinear
N{\'e}el-like ferrimagnetic component exhibiting long-range order
below $T_C$ and a spiral component, which exhibits only short-range
order even in the lowest temperature phase.

In order to assess the validity of the LKDM theory and to facilitate a
better comparison with experimental data, an independent determination
of the magnetic coupling constants in these systems is very
desirable. Density functional theory (DFT, see
Ref.~\onlinecite{Jones/Gunnarsson:1989}) provides an efficient way for
the {\it ab initio} determination of such magnetic coupling constants
that can then be used for an accurate modeling of the spin structure
of a particular system. DFT also offers a straightforward way to
investigate the effect of structural distortions on the magnetic
coupling constants, and is therefore ideally suited to study the
coupling between magnetism and structural properties.

Traditionally, insulating magnetic oxides represent a great challenge
for DFT-based methods due to the strong Coulomb interaction between
the localized $d$ electrons. However, recently the local spin density
approximation plus Hubbard $U$ (LSDA+$U$) method has been very
successful in correctly determining various properties of such
strongly correlated magnetic
insulators.\cite{Anisimov/Aryatesiawan/Liechtenstein:1997} In
particular, it has been used for the calculation of magnetic coupling
constants in a variety of transition metal
oxides.\cite{Solovyev/Terakura:1998,Yaresko_et_al:2000,Baettig/Ederer/Spaldin:2005,Novak/Rusz:2005,Fennie/Rabe:2006}

Here we present an LSDA+$U$ study of the magnetic coupling constants
in the spinel systems CoCr$_2$O$_4$ and MnCr$_2$O$_4$. The goal of the
present paper is to provide accurate values for the relevant coupling
constants in these two systems, in order to test the assumptions made
by LKDM and to resolve the uncertainties in the interpretation of the
experimental data. In addition, we assess the general question of how
accurate such magnetic coupling constants in complex oxides can be
determined using the LSDA+$U$ method.

We find that in contrast to the assumptions of the LKDM theory, the
coupling between the $A$ site cations is not necessarily negligible,
but that the general validity of the LKDM theory should be better for
CoCr$_2$O$_4$ than for MnCr$_2$O$_4$, in agreement with what has been
concluded from the experimental data. However, in contrast to what
follows from fitting the experimental data to the LKDM theory, the
calculated $u$ for CoCr$_2$O$_4$ is smaller than the value of $u=2.0$
obtained from the experimental fit. In addition, we show that by
analyzing the dependence of the magnetic coupling constants on the
LSDA+$U$ parameters and on the lattice constant, the various
mechanisms contributing to the magnetic interaction can be identified,
and a quantitative estimate of the corresponding coupling constant can
be obtained within certain limits.

The present paper is organized as follows. In Sec.~\ref{methods} we
present the methods we use for our calculations. In particular, we
give a brief overview over the LSDA+$U$ method and the challenges in
using this method as a quantitative and predictive tool. In
Sec.~\ref{sec:results} we present our results for the lattice
parameters, electronic structure, and magnetic coupling constants of
the two investigated Cr spinels. Furthermore, we analyze in detail the
dependence of the magnetic coupling constants on the lattice constant
and LSDA+$U$ parameters, and we discuss the reasons for the observed
trends. We end with a summary of our main conclusions.

\section{Methods}
\label{methods}

\subsection{LSDA+$U$}
\label{sec:methods}

The LSDA+$U$ method offers an efficient way to calculate the
electronic and magnetic properties of complex transition metal
oxides. The idea behind the LSDA+$U$ method is to explicitly include
the Coulomb interaction between strongly localized $d$ or $f$
electrons in the spirit of a mean-field Hubbard model, whereas the
interactions between the less localized $s$ and $p$ electrons are
treated within the standard local spin density approximation
(LSDA).\cite{Anisimov/Zaanen/Andersen:1991} To achieve this, a
Hubbard-like interaction term $E_U$, which depends on the occupation
of the localized orbitals, is added to the LSDA total energy, and an
additional double counting correction $E_\text{dc}$ is introduced to
subtract that part of the electron-electron interaction between the
localized orbitals that is already included in the LSDA:
\begin{equation}
\label{eq:ldau}
E = E_\text{LSDA} + E_U - E_\text{dc} \quad.
\end{equation} 
Here
\begin{equation}
\label{E_U}
E_U = \frac{1}{2} \sum_{ \{ \gamma \} } \left( U_{\gamma_1 \gamma_3 \gamma_2
  \gamma_4} - U_{\gamma_1 \gamma_3 \gamma_4 \gamma_2} \right) n_{\gamma_1
  \gamma_2} n_{\gamma_3 \gamma_4}
\end{equation}
and
\begin{equation}
\label{eq:edc}
E_\text{dc} = \frac{U}{2} n (n-1) - \frac{{\cal J}^H}{2} \sum_s n_s
(n_s - 1) \quad ,
\end{equation}
where $\gamma = (m,s)$ is a combined orbital and spin index of the
correlated orbitals, $n_{\gamma_1\gamma_2}$ is the corresponding
orbital occupancy matrix, $n_s = \sum_m n_{ms,ms}$ and $n = \sum_s
n_s$ are the corresponding traces with respect to spin and both spin
and orbital degrees of freedom, and $U_{\gamma_1\gamma_3
\gamma_2\gamma_4} = \langle m_1m_3|V_\text{ee}|m_2m_4 \rangle
\delta_{s_1s_2}\delta_{s_3s_4}$ are the matrix elements of the
screened electron electron interaction, which are expressed as usual
in terms of two parameters, the Hubbard $U$ and the intra-atomic
Hund's rule parameter ${\cal J}^H$ (see
Ref.~\onlinecite{Anisimov/Aryatesiawan/Liechtenstein:1997}).

The LSDA+$U$ method has been shown to give the correct ground states
for many strongly correlated magnetic insulators, and thus represents
a significant improvement over the LSDA for such
systems.\cite{Anisimov/Aryatesiawan/Liechtenstein:1997} Furthermore,
the LSDA+$U$ method is very attractive due to its simplicity and the
negligible additional computational effort compared to a conventional
LSDA calculation. It therefore has become a widely used tool for the
study of strongly correlated magnetic insulators. Since the LSDA+$U$
method treats the interactions between the occupied orbitals only in
an effective mean-field way, it fails to describe systems where
dynamic fluctuations are important. For such systems, the local
density approximation plus dynamical mean field theory (LDA+DMFT),
which also includes local dynamic correlations, has been introduced
recently.\cite{Anisimov_et_al:1997} However, the LDA+DMFT method is
computationally rather demanding, and is currently too costly to be
used for the calculation of magnetic characteristics of such complex
materials as the spinels. On the other hand, for a large number of
systems such fluctuations are only of minor importance, and for these
systems the LSDA+$U$ method leads to a good description of the
electronic and magnetic properties.

However, in order to obtain reliable results, the use of the LSDA+$U$
method should be accompanied by a careful analysis of all the
uncertainties inherent in this method. An additional goal of the
present paper is therefore to critically assess the predictive
capabilities of the LSDA+$U$ method for the determination of magnetic
coupling constants in complex transition metal oxides. Apart from the
question about the general applicability of the LSDA+$U$ approach to
the investigated system, and the unavoidable ambiguities in the
definition of the LSDA+$U$ energy functional
(Eqs.~(\ref{eq:ldau})-(\ref{eq:edc})),\cite{Solovyev/Dederichs/Anisimov:1994,Czyzyk/Sawatzky:1994}
the proper choice of the parameters $U$ and ${\cal J}^H$ represents
one of the main hurdles when the LSDA+$U$ method is used as a
quantitative and predictive tool.

$U$ and ${\cal J}^H$ can in principle be calculated using constrained
density functional theory,\cite{Dederichs_et_al:1984} thus rendering
the LSDA+$U$ method effectively parameter-free. In practice however,
the exact definition of $U$ and ${\cal J}^H$ within a solid is not
obvious, and the calculated values depend on the choice of orbitals or
the details of the method used for their
determination.\cite{Hybertsen/Schlueter/Christensen:1989,Solovyev/Hamada/Terakura:1996,Pickett/Erwin/Ethridge:1998,Cococcioni/Gironcoli:2005}
Therefore, parameters obtained for a certain choice of orbitals are
not necessarily accurate for calculations using a different set of
orbitals.

In the present work we thus pursue a different approach. We choose
values for $U$ and ${\cal J}^H$ based on a combination of previous
constrained DFT calculations, experimental data, and physical
reasoning, and these values are then varied within reasonable limits
to study the resulting effect on the physical properties. In
particular, for the spinel systems studied in this work the Hubbard
$U$s on the transition metal sites are varied between 2\,eV and 6\,eV
(in 1\,eV increments), with the additional requirement that
$U_\text{Cr} \leq U_{A}$ ($A$ = Co, Mn). For the on-site Hund's rule
coupling we use two different values, ${\cal J}^H$ = 0\,eV and ${\cal
J}^H$ = 1\,eV with ${\cal J}^H_\text{Cr} = {\cal J}^H_{A}$. The
conditions $U_\text{Cr} \leq U_{A}$ and ${\cal J}^H_\text{Cr} = {\cal
J}^H_{A}$ are motivated by constrained DFT calculations for a series
of transition metal perovskite systems, which showed that the Hubbard
$U$ increases continuously from V to Cu, whereas the on-site exchange
parameter ${\cal J}^H$ is more or less constant across the
series.\cite{Solovyev/Hamada/Terakura:1996} A similar trend for $U$
can be observed in the simple transition metal
monoxides.\cite{Anisimov/Zaanen/Andersen:1991,Pickett/Erwin/Ethridge:1998}
Although in the spinel structure the coordination and formal charge
state of the $A$ cation is different from the $B$ cation, we assume
that the assumption $U_\text{Cr} \leq U_{A}$ is nevertheless valid,
since the screening on the sixfold coordinated $B$ site is expected to
be more effective than on the tetrahedral $A$ site. Further evidence
for the validity of this assumption is given by the relative widths of
the $d$ bands on the $A$ and $B$ sites obtained from the calculated
orbitally resolved densities of states (see Fig.~\ref{fig:dos} and
Sec.~\ref{DOS}).

The absolute values of $U$ used in this work are motivated by recent
constrained DFT calculations using linear response
techniques,\cite{Pickett/Erwin/Ethridge:1998,Cococcioni/Gironcoli:2005}
which lead to significantly smaller values of $U$ than previous
calculations using the linear muffin tin orbital (LMTO) method, where
the occupation numbers are constrained by simply setting all transfer
matrix elements out of the corresponding orbitals to
zero.\cite{Anisimov/Gunnarsson:1991,Solovyev/Hamada/Terakura:1996}
Typical values obtained for various transition metal ions in different
chemical environments are between
3-6\,eV.\cite{Pickett/Erwin/Ethridge:1998,Cococcioni/Gironcoli:2005}
For the Cr$^{3+}$ ion a value of $U \approx 3$\,eV, derived by
comparing the calculated densities of states with photo-emission data,
has been used
successfully.\cite{Fennie/Rabe_CCS:2005,Fennie/Rabe:2006} We thus
consider the values $U_A$ = 4-5\,eV and $U_\text{Cr}$ = 3\,eV as the
most adequate $U$ parameters for our systems. Nevertheless, we vary
these parameters here over a much larger range, in order to see and
discuss the resulting trends in the calculated magnetic coupling
constants.

For the Hund's rule parameter ${\cal J}^H$ screening effects are less
important, and calculated values for various systems are all around or
slightly lower than
1\,eV.\cite{Anisimov/Zaanen/Andersen:1991,Solovyev/Hamada/Terakura:1996}
On the other hand, a simplified LSDA+$U$ formalism is sometimes used,
where the only effect of ${\cal J}^H$ is to reduce the effective
Coulomb interaction $U_\text{eff} = U - {\cal
J}^H$.\cite{Sawada_et_al:1997,Dudarev_et_al:1998,Ederer/Spaldin_2:2005}
In this work we use the two values ${\cal J}^H$ = 0\,eV and ${\cal
J}^H$ = 1\,eV to study the resulting effect on the magnetic coupling
constants.

\subsection{Other technical details}
\label{sec:details}

To determine the magnetic coupling constants corresponding to the
closest neighbor magnetic interactions between the various
sublattices, we calculate the total energy differences for four
different collinear magnetic configurations: the N{\'e}el type
ferrimagnetic order, the ferromagnetic configuration, and two
different configurations with anti-parallel magnetic moments within
the $A$ and $B$ sub-lattices respectively, and we then project the
resulting total energies on a simple classical Heisenberg model,
\begin{equation}
E = - \sum_{i,j} \tilde{J}_{ij} \vec{S}_i \cdot \vec{S}_j = -
\sum_{i,j} J_{ij} \hat{e}_i \cdot \hat{e}_j \quad ,
\label{eq:Heisenberg}
\end{equation}
where only the nearest neighbor coupling constants $J_{AB}$, $J_{BB}$,
and $J_{AA}$ are assumed to be nonzero, and where we defined the
coupling constants $J_{ij} = \tilde{J}_{ij} S_i S_j$ corresponding to
normalized spin directions $\hat{e}_i$ of the magnetic ions. We note
that even though for itinerant systems such as the elementary magnets
Fe, Co, and Ni, the coupling constants obtained in this way can be
different from the ones obtained for only small variations from the
collinear configurations,\cite{Liechtenstein/Katsnelson/Gubanov:1984}
the local magnetic moments of many insulating transition metal oxides,
in particular the systems investigated in the present study, behave
much more like classical Heisenberg spins and thus the simpler
approach pursued in this work is justified. We point out that a
determination of all possible further neighbor interactions is beyond
the scope of this paper and is therefore left for future studies.

We perform calculations at both experimentally determined lattice
constants and theoretical lattice parameters. The theoretical lattice
parameters are obtained by a full structural relaxation within the LSDA
for a collinear N{\'e}el-type magnetic configuration. The same LSDA
lattice parameters are used in all our calculations with varying
values of the LSDA+$U$ parameters $U$ and ${\cal J}^H$. In order to
reduce the required computational effort, we do not perform
relaxations for each individual set of LSDA+$U$ parameters. Except
when noted otherwise, all calculations are performed using the
``Vienna Ab-initio Simulation Package'' (VASP) employing the projector
augmented wave (PAW)
method.\cite{Bloechl:1994,Kresse/Furthmueller_PRB:1996,Kresse/Joubert:1999}
We use a plane wave energy cutoff of 450\,eV (550\,eV for relaxations)
and a 5$\times$5$\times$5 $\Gamma$-centered mesh for Brillouin zone
integrations.  Increasing the mesh density by using a
8$\times$8$\times$8 mesh results only in negligible changes for the
calculated total energy differences. Structural relaxations are
performed until the forces are less than 10$^{-5}$ eV/\AA\ and all
components of the stress tensor are smaller than 0.02 kbar. The
electronic self-consistency cycle is iterated until the total energy
is converged better than 10$^{-8}$\,eV. In addition, we perform some
test calculations using the full-potential linear-augmented-plane-wave
(FLAPW) method.\cite{Wimmer:1981} For these calculations we use the
Wien97 code\cite{Blaha:1990} with our own implementation of the
LSDA+$U$ method.  The plane-wave cut-off parameter is set to $223$\,eV
in these calculation, and the Brillouin-zone integration is also
carried out on a 5$\times$5$\times$5 $\Gamma$-centered mesh. The
criterion for self-consistency is the difference in the total energy
after the last two iterations being less than $10^{-4}$\,Ry.

\section{Results and discussion}
\label{sec:results}

\subsection{Structural relaxation}

\begin{table}
\caption{Structural parameters calculated in this work. $a$ is the
 lattice constant of the cubic spinel structure, and the internal
 structural parameter $x$ corresponds to the Wyckoff position 32e
 ($x$,$x$,$x$) of the oxygen sites. Columns ``theo.'' contain the
 values calculated in this work while columns ``exp.'' contain
 experimental data.}
\label{tab:struc}
\begin{ruledtabular}
\begin{tabular}{c|cc|cc}
 & \multicolumn{2}{c|}{CoCr$_2$O$_4$} & \multicolumn{2}{c}{MnCr$_2$O$_4$} \\
 & exp. (Ref.~\onlinecite{Lawes_et_al:2006}) & theo. &
 exp. (Ref.~\onlinecite{Ram_private}) & theo. \\
\hline
$a$ [\AA] & 8.335 & 8.137 & 8.435 & 8.242 \\
$x$ & 0.264 & 0.260 & 0.264 & 0.262 \\
\end{tabular}
\end{ruledtabular}
\end{table}

Table~\ref{tab:struc} shows the structural parameters obtained in this
work together with corresponding experimental data. The theoretical
lattice constants are obtained within the LSDA and for N{\'e}el-type
ferrimagnetic order, and are about 2.3~\% smaller than the
corresponding experimental values for both materials. The calculated
internal structural parameters $x$ are in very good agreement with
experiment. The underestimation of the lattice constant by a few
percent is a typical feature of the LSDA in complex transition metal
oxides.\cite{Neaton_et_al:2005,Fennie/Rabe:2006}

\subsection{Electronic structure}
\label{DOS}

\begin{figure}[htbp]
\centerline{\includegraphics*[width=0.9\columnwidth]{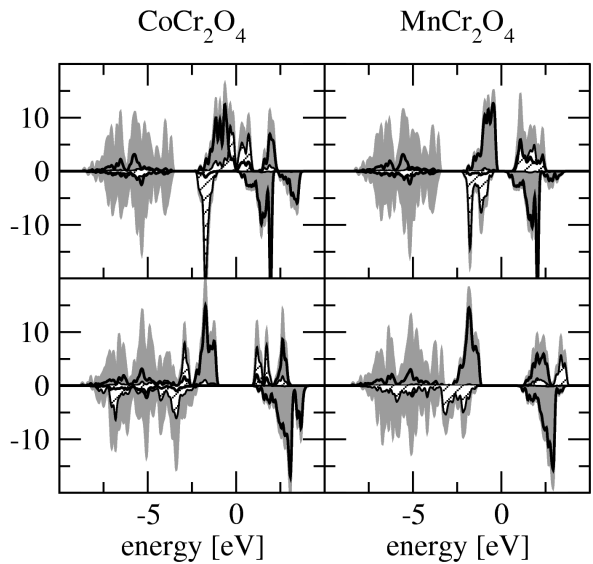}}
\caption{Densities of states (in states/eV) for CoCr$_2$O$_4$ (left
  two panels) and MnCr$_2$O$_4$ (right two panels) calculated using
  the LSDA (upper two panels) and the LSDA+$U$ method with $U_A$ =
  5\,eV, $U_\text{Cr}$ = 3\,eV, and ${\cal J}^H_A$ = ${\cal
  J}^H_\text{Cr}$ = 0\,eV (lower two panels). Calculations were done
  at the experimental lattice parameters for a collinear N{\'e}el-type
  ferrimagnetic structure where the direction of the Cr magnetic
  moments was defined as ``spin-up'', and corresponding ``spin-down''
  states are shown with a negative sign. The gray shaded areas
  represent the total density of states, the curves shaded with
  diagonal lines represent the $d$ states on the $A$ site of the
  spinel lattice, and the thick black lines correspond to the Cr $d$
  states. Zero energy separates the occupied from the unoccupied
  states.}
\label{fig:dos}
\end{figure}

Fig.~\ref{fig:dos} shows the densities of states for both
CoCr$_2$O$_4$ and MnCr$_2$O$_4$ calculated using the LSDA and the
LSDA+$U$ method with $U_A$ = 5\,eV, $U_\text{Cr}$ = 3\,eV, ${\cal
J}^H_A$ = ${\cal J}^H_\text{Cr}$ = 0\,eV, and a collinear
N{\'e}el-type magnetic configuration at the experimental lattice
constants. Both systems are insulating within the LSDA, although the
LSDA energy gap for CoCr$_2$O$_4$ is very small, about 0.15\,eV. The
LSDA gap is larger for MnCr$_2$O$_4$, since in this system the gap is
determined by the relatively strong crystal-field splitting on the
octahedral $B$ site and the equally strong magnetic splitting, whereas
in CoCr$_2$O$_4$ the width of the LSDA gap is limited by the small
crystal-field splitting on the tetrahedrally coordinated $A$ site. The
use of the LSDA+$U$ method increases the width of the energy gap
substantially and pushes the majority $d$ states on the $A$ site down
in energy, leading to strong overlap with the oxygen 2$p$ states. In
the LSDA the transition metal $d$ states are well separated from the
oxygen $p$ manifold, whereas the LSDA+$U$ method increases the
energetic overlap between these states. In all cases the gap is
between occupied and unoccupied transition metal $d$ states.

It can be seen that the bandwidth of the $d$-bands for the
tetrahedrally coordinated $A$ site is indeed smaller than for the
octahedral $B$ site. Thus, the $d$ states on the $A$ sites are more
localized and one can expect a larger on-site Coulomb interaction than
on the Cr $B$ site, in agreement with the assumption that $U_\text{Cr}
\leq U_{A}$ (see Sec.~\ref{sec:methods}).

\subsection{Magnetic coupling constants}

\begin{figure}
\centerline{\includegraphics*[width=0.9\columnwidth]{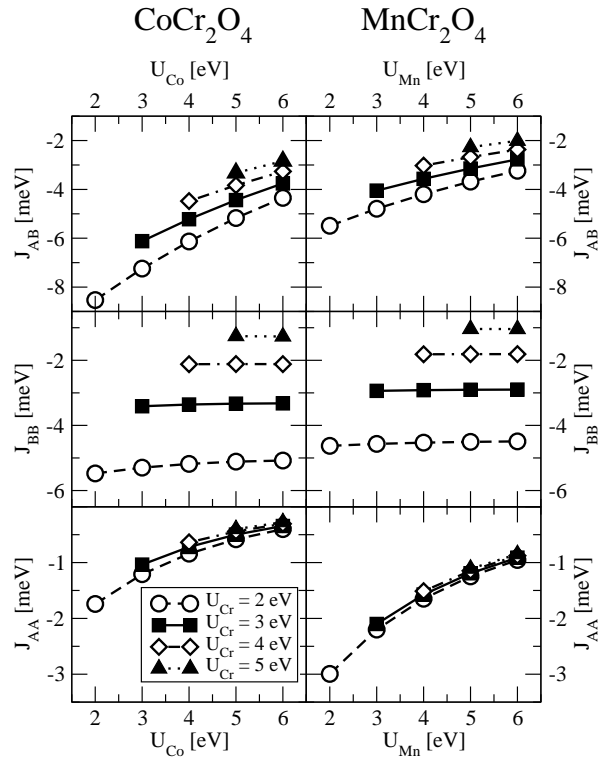}}
\caption{Magnetic coupling constants $J_{AB}$ (upper panels), $J_{BB}$
  (middle panels), and $J_{AA}$ (lower panels) calculated for
  CoCr$_2$O$_4$ (left) and MnCr$_2$O$_4$ (right) as a function of
  $U_A$ for $U_\text{Cr}$ = 2\,eV (open circles), 3\,eV (filled
  squares), 4\,eV (open diamonds), and 5\,eV (filled triangles). All
  calculations were performed using experimental lattice parameters
  and ${\cal J}^H_{A}$ = ${\cal J}^H_\text{Cr}$ = 1\,eV.}
\label{fig:cco1}
\end{figure}

Fig.~\ref{fig:cco1} shows the magnetic coupling constants calculated
using the experimental lattice parameters, ${\cal J}^H_{A} = {\cal
J}^H_\text{Cr} = 1$\,eV, and different values of the Hubbard $U$ on
the $A$ and $B$ sites. All coupling constants are negative,
i.e. antiferromagnetic, and decrease in strength when the Hubbard
parameters are increased. The ``inter-sublattice'' coupling $J_{AB}$
depends similarly on both $U_A$ and $U_B$, whereas both
``intra-sublattice'' coupling constants $J_{BB}$ and $J_{AA}$ depend
only on the Hubbard parameter of the corresponding sublattice.

The $BB$ interaction in the spinel lattice is known to result from a
competition between antiferromagnetic (AFM) direct cation-cation
exchange and indirect cation-anion-cation exchange, which for the
present case of a 90$^\circ$ cation-anion-cation bond angle gives rise
to a ferromagnetic (FM) interaction.\cite{Goodenough:Book} The AFM
direct interaction is expected to dominate at smaller volumes, whereas
at larger volumes the FM indirect interaction should be
stronger. Furthermore, it is important to note that even the pure
direct cation-cation interaction is comprised out of two parts: (i)
the ``potential exchange'' due to the standard Heitler-London
exchange-integral, which is always FM for orthogonal orbitals but is
usually negligible, and (ii) the AFM ``kinetic exchange'', which
results from a second order perturbation treatment of the electron
hopping and is proportional to
1/$U$.\cite{Anderson:1963,Goodenough:Book} The observed $U$ dependence
of $J_{BB}$ can thus be understood as follows: At small values of $U$
the AFM direct kinetic exchange is strongest, but it is suppressed as
the value of $U$ is increased. The FM indirect cation-anion-cation
exchange also decreases, but in addition increasing $U$ shifts the
cation $d$ states down in energy and thus leads to enhanced
hybridization with the anion $p$ states (see Sec.~\ref{DOS}). This
enhanced hybridization partially compensates the effect of increasing
$U$ so that the indirect exchange decreases slower than 1/$U$.
Therefore, the FM indirect exchange is less affected by increasing $U$
than the AFM direct exchange, and thus gains in strength relative to
the latter.  This explains why the observed decrease of $J_{BB}$ is
stronger than $1/U$.  In fact, for the larger experimental volumes and
using $J = 0$\,eV for the Hund's rule coupling (see discussion below)
the $BB$ coupling in both systems even becomes slightly FM for large
$U$.

The $AB$ coupling in the spinels is mediated by a cation-anion-cation bond
with an intermediate angle of $\sim$120$^\circ$, which makes it difficult to
predict the sign of the coupling based on general considerations. A weak AFM
interaction has been proposed for the case of empty $e_g$ orbitals on the $B$
site,\cite{Wickham/Goodenough:1959} in agreement with the present results.

Comparing the values of $J_{AB}$ and $J_{BB}$ calculated for a
constant set of LSDA+$U$ parameters shows that both are of the same
order of magnitude. On the other hand $J_{AA}$ is expected to be
significantly weaker, since it corresponds to a cation-cation distance
of $\sim 3.6$\,\AA\ with the shortest superexchange path along an
$A$-O-O-$A$ bond sequence. Based on this assumption $J_{AA}$ was
neglected in the theoretical treatment of LKDM.\cite{Lyons_et_al:1962}
Nevertheless, in our calculations $J_{AA}$ is found to be of
appreciable strength.  This is particular striking for MnCr$_2$O$_4$,
but also for CoCr$_2$O$_4$ the difference between $J_{AA}$ and
$J_{AB}$ ($J_{BB}$) is less than an order of magnitude. From this it
follows that $J_{AA}$ is definitely non-negligible in MnCr$_2$O$_4$
and can also lead to significant deviations from the LKDM theory in
CoCr$_2$O$_4$. We point out that this conclusion holds true
independently of the precise values of the LSDA+$U$ parameters used in
the calculation. An appreciable value for $J_{AA}$ has also been found
in a previous LSDA study of the spinel ferrite
MnFe$_2$O$_4$.\cite{Singh/Gupta/Gupta:2002}

As stated in Sec.~\ref{sec:details}, a full determination of all
possible further neighbor interactions is beyond the scope of this
paper. However, to obtain a rough estimate of the strength of further
neighbor interactions in CoCr$_2$O$_4$, and to see whether this
affects the values of $J_{AB}$, $J_{BB}$, and $J_{AB}$ obtained in
this work, we perform some additional calculations using a doubled
unit cell. This allows us to determine the coupling constant
$J^{(3)}_{BB}$, corresponding to the third nearest neighbor $BB$
coupling. As shown in Ref.~\onlinecite{Dwight/Menyuk:1967}, due to the
special geometry of the spinel structure, this third nearest neighbor
coupling is larger than all other further neighbor interactions within
the $B$ sublattice, and can be expected to represent the next
strongest magnetic interaction apart from $J_{BB}$, $J_{AB}$, and
$J_{AA}$. This coupling is mediated by a $B$-O-O-$B$ bond sequence and
corresponds to a $BB$ distance of 5.89\,\AA. For comparison, the
distances corresponding to $J_{BB}$, $J_{AB}$, and $J_{AA}$ are
2.94\,\AA, 3.46\,\AA, and 3.61\,\AA, respectively.\cite{distances} For
these test calculations we use the experimental lattice parameters of
CoCr$_2$O$_4$ and the LSDA+$U$ parameters $U_\text{Co}=5$\,eV,
$U_\text{Cr}=3$\,eV, and ${\cal J}^H=0$\,eV. We obtain a value of
$J^{(3)}_{BB} = 0.15$\,eV, corresponding to a weak FM
coupling. However, the magnitude of $J^{(3)}_{BB}$ is small compared
to $J_{AB}$ and $J_{BB}$, and we therefore continue to neglect further
neigbor interactions in the following.

Next we calculate the magnetic coupling constants using the lattice
parameters obtained by a full structural optimization within the LSDA,
and also by using ${\cal J}^H = 0$\,eV at both experimental and
theoretical lattice parameters. Again, we vary $U_{A}$ and
$U_\text{Cr}$ independently. The observed variation of the coupling
constants with respect to the Hubbard parameters is very similar to
the case shown in Fig.~\ref{fig:cco1}, only the overall magnitude of
the magnetic coupling constants is changed. We therefore discuss only
the results obtained for $U_\text{Cr} =3$\,eV and $U_{A} = 5$\,eV,
which are physically reasonable choices for these parameters, as
discussed in Sec.~\ref{sec:methods}.

\begin{table}
\caption{Calculated magnetic coupling constants $J_{AB}$, $J_{BB}$,
  and $J_{AA}$ for different lattice parameters and different values
  of the intra-atomic Hund's rule coupling parameter ${\cal J}^H$ for
  $U_A$ = 5\,eV and $U_\text{Cr}$ = 3\,eV. Lattice parameters ``exp.''
  and ``theo.''  refer to the corresponding values listed in
  Table~\ref{tab:struc}.}
\label{tab:coupling}
\begin{ruledtabular}
\begin{tabular}{cc|cccc}
& ${\cal J}^H$ (eV) & 1.0 & 0.0 & 1.0 & 0.0 \\
& latt. param. & exp. & exp. & theo. & theo. \\
\hline
& $J_{AB}$ (meV) & $-$4.44 & $-$3.55 & $-$6.02 & $-$4.83 \\
CoCr$_2$O$_4$ & $J_{BB}$ (meV) & $-$3.33 & $-$1.04 & $-$6.90 & $-$4.34 \\
& $J_{AA}$ (meV) & $-$0.50 & $-$0.44 & $-$0.77 & $-$0.58 \\
\hline
& $J_{AB}$ (meV) & $-$3.14 & $-$1.40 & $-$4.88 & $-$2.61 \\
MnCr$_2$O$_4$ & $J_{BB}$ (meV) & $-$2.91 & $-$0.74 & $-$5.22 & $-$2.74 \\
& $J_{AA}$ (meV) & $-$1.19 & $-$0.92 & $-$1.88 & $-$1.45 \\
\end{tabular}
\end{ruledtabular}
\end{table}

Table~\ref{tab:coupling} shows the calculated magnetic coupling
constants for the different cases. It is apparent that both volume and
the intra-atomic exchange parameter ${\cal J}^H$ have a significant
effect on the calculated results.  The volume dependence can easily be
understood. The smaller theoretical volume leads to stronger coupling
between the magnetic ions. This is particularly significant for
${J}_{BB}$, since the direct exchange interaction between the $B$
cations is especially sensitive to the inter-cation distance. The
corresponding coupling is therefore strongly enhanced (suppressed) by
decreasing (increasing) the lattice constant.  The indirect
superexchange interaction also depends strongly on the inter-atomic
distances.

It can be seen from Table~\ref{tab:coupling} that ${\cal J}^H=0$
significantly decreases the strength of all magnetic coupling
constants compared to ${\cal J}^H=1$\,eV. A strong ${\cal J}^H$
dependence of the magnetic coupling has also been observed in other Cr
spinels with non-magnetic cations on the $A$
site.\cite{Craig:unpublished} Further calculations, with different
values for ${\cal J}^H$ on the $A$ and $B$ sites respectively, show
that it is mostly ${\cal J}^H_\text{Cr}$ which is responsible for this
effect. On the other hand, varying ${\cal J}^H_{A}$ has a smaller
effect on the magnetic coupling. This is consistent with the very
strong ${\cal J}^H$ dependence of $J_{BB}$ and the weaker ${\cal J}^H$
dependence of $J_{AA}$ seen in Table~\ref{tab:coupling}.

To understand the strong effect of ${\cal J}^H_\text{Cr}$ on the
magnetic coupling constants we first take a look at the occupation
numbers $n_{\gamma} \equiv n_{\gamma\gamma}$ of the Cr $d$
orbitals. The corresponding occupation numbers in CoCr$_2$O$_4$ are
(calculated for a FM configuration at the experimental lattice
parameters and using ${\cal J}^H = 0$): $n_{t_{2g},\uparrow} = 0.95$,
$n_{t_{2g},\downarrow} = 0.05$, $n_{e_{g},\uparrow} = 0.32$, and
$n_{e_{g},\downarrow} = 0.21$. As expected, the occupation of the
$t_{2g}$ orbitals represents the formal $d^3$ valency with full
spin-polarization, but in addition there is a sizable $e_g$
occupation, which contributes $\sim 0.2 \mu_\text{B}$ to the local
spin moment of the Cr cation. This partial $e_g$ occupation, which is
due to hybridization with the oxygen $p$ bands, gives rise to a FM
interaction between the Cr sites, because the $e_g$ polarization is
coupled to the $t_{2g}$ spins via the Hund's rule
coupling.\cite{Goodenough:Book} This FM interaction between the Cr
sites should therefore be proportional to the strength of the Hund's
rule coupling. Thus, the stronger AFM interaction for ${\cal
J}^H=1$\,eV compared to ${\cal J}^H=0$ (see Table~\ref{tab:coupling})
might be surprising at first. However, it is important to realize that
even though the parameter ${\cal J}^H$ represents the strength of the
Hund's rule coupling, its effect within the LSDA+$U$ framework is not
to introduce a strong Hund's rule interaction. If one analyzes the
LSDA+$U$ energy expression, Eq.~(\ref{eq:ldau}), in a somewhat
simplified picture where the occupation matrix is diagonal and the
Coulomb matrix elements are orbitally independent, one can see that
the double counting correction, $E_\text{dc}$, exactly cancels the
different potential shifts for orbitals with parallel and antiparallel
spins that are caused by $E_U$ for ${\cal J}^H \neq 0$, if one of the
$d$ orbitals is filled. Thus, $E_U-E_\text{dc}$ does not lead to an
additional Hund's rule interaction compared to $E_\text{LSDA}$, even
for ${\cal J}^H \neq 0$. It is generally assumed that this type of
interaction is already well described on the LSDA level. The only
effect of ${\cal J}^H$ is therefore an effective reduction of the
on-site Coulomb repulsion. This can be seen in the following, where we
write the simplified version of Eq.~(\ref{eq:ldau}) as (see
Ref.~\onlinecite{Sawada_et_al:1997}):
\begin{equation}
\label{eq:ldau2}
E = E_\text{LSDA} + \frac{U_\text{eff}}{2} \left( n - \sum_{\gamma}
n_\gamma n_\gamma \right) \quad ,
\end{equation}
with $U_\text{eff} = U -{\cal J}^H$. Within this simplified LSDA+$U$
version, the effect of ${\cal J}^H$ on the magnetic coupling constant
can be understood as an effective reduction of the on-site Coulomb
interaction. According to the previously discussed $U$ dependence of
the magnetic coupling constants (see also Fig.~\ref{fig:cco1}), a
reduced on-site Coulomb interaction leads to a stronger AFM
interaction for all calculated magnetic coupling constants.

From Fig.~\ref{fig:cco1} it can be seen that the magnetic coupling
constants for CoCrO$_2$ using experimental lattice parameters and the
LSDA+$U$ parameters $U_\text{Co}=6$\,eV, $U_\text{Cr}=4$\,eV, and
${\cal J}^H=1$\,eV, i.e. $U_\text{eff,Co}=5$\,eV and
$U_\text{eff,Cr}=3$\,eV, are: $J_{AB}=-3.26$\,eV, $J_{BB}=-2.12$\,eV,
and $J_{AA}=-0.30$\,eV. The corresponding result for
$U_\text{Co}=5$\,eV, $U_\text{Cr}=3$\,eV, and ${\cal J}^H=0$, i.e. for
the same values of $U_\text{eff}$ but different ${\cal J}^H$, are:
$J_{AB}=-3.55$\,eV, $J_{BB}=-1.04$\,eV, and $J_{AA}=-0.44$\,eV (see
Table~\ref{tab:coupling}). Thus, the pure dependence on $U_\text{eff}$
seems to be approximately valid for $J_{AB}$ and $J_{AA}$, whereas
there is a notable quantitative deviation from the simplified LSDA+$U$
model in the case of $J_{BB}$. Nevertheless, the overall trend can
still be understood from the simplified LSDA+$U$ pictue.

\begin{table}
\caption{Magnetic coupling constants of CoCr$_2$O$_4$ and
MnCr$_2$O$_4$ calculated using two different methods (FLAPW and PAW),
different values for ${\cal J}^H$, and the experimental lattice
parameters.}
\label{tab:compare}
\begin{ruledtabular}
\begin{tabular}{llcccc}
& & \multicolumn{2}{c}{FLAPW} & \multicolumn{2}{c}{PAW} \\
              & ${\cal J}^H$ (eV) & 0.0 & 1.0 & 0.0 & 1.0 \\
\hline
              & $J_{AB}$ (meV) & $-$3.62 & $-$4.32 & $-$3.55 & $-$4.44 \\
CoCr$_2$O$_4$ & $J_{BB}$ (meV) & $-$1.32 & $-$3.09 & $-$1.04 & $-$3.33 \\
              & $J_{AA}$ (meV) & $-$0.23 & $-$0.00 & $-$0.44 & $-$0.50 \\
\hline
              & $J_{AB}$ (meV) & $-$1.73 & $-$3.23 & $-$1.40 & $-$3.14 \\
MnCr$_2$O$_4$ & $J_{BB}$ (meV) & $-$1.32 & $-$3.21 & $-$0.74 & $-$2.91 \\
              & $J_{AA}$ (meV) & $-$0.67 & $-$1.06 & $-$0.92 & $-$1.19  
\end{tabular}
\end{ruledtabular}
\end{table}

Finally, to assess the possible influence of different methods to
solve the self-consistent Kohn-Sham equations on the calculated
magnetic coupling constants, we perform additional tests using a
different electronic structure code employing the FLAPW method (see
Sec.~\ref{sec:details}). The results are summarized and compared to
the PAW results in Table~\ref{tab:compare}. There are some variations
in the absolute values of the magnetic coupling constants obtained
with the two different methods, but overall the agreement is rather
good. Trends are the same in both methods, and in particular the
strong effect of the LSDA+$U$ Hund's rule parameter ${\cal J}^H$ on
the magnetic coupling constants is confirmed by the FLAPW
calculations. One possible reason for the differences between the PAW
results and the FLAPW results is that the radii of the projection
spheres used in the PAW method are chosen differently from the radii
of the Muffin-Tin spheres used to construct the FLAPW basis functions.

\subsection{The LKDM parameter $u$}

\begin{figure}
\centerline{\includegraphics*[width=0.8\columnwidth]{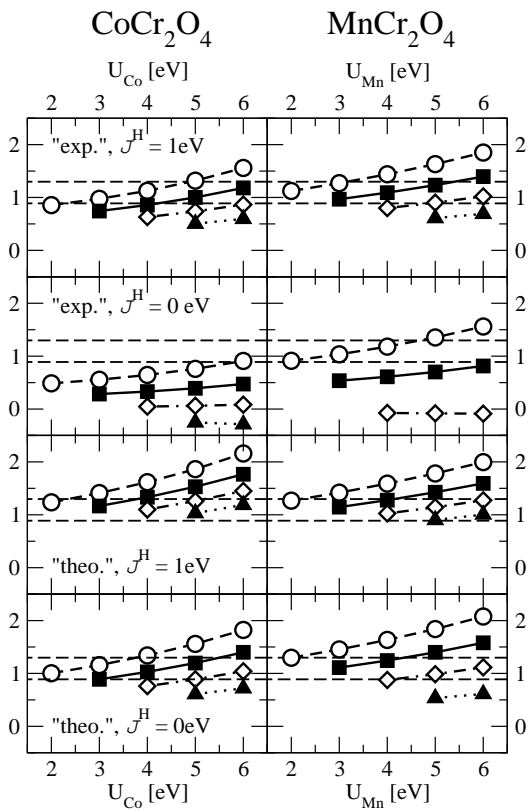}}
\caption{Dependence of the LKDM parameter $u$ on the Hubbard $U$
  parameters of the different magnetic cations. Left panels correspond
  to CoCr$_2$O$_4$, right panels to MnCr$_2$O$_4$. From top to bottom
  the different panels correspond to calculations for exp. volume and
  ${\cal J}^H = 1$\,eV, exp. volume and ${\cal J}^H = 0$\,eV,
  theo. volume and ${\cal J}^H = 1$\,eV, and theo. volume and ${\cal
  J}^H = 0$\,eV (open circles: $U_\text{Cr} = 2$\,eV, filled squares:
  $U_\text{Cr} = 3$\,eV, open diamonds: $U_\text{Cr} = 4$\,eV, filled
  triangles: $U_\text{Cr} = 5$\,eV).  Dashed horizontal lines indicate
  the critical values $u_0 = 8/9$ and $u''\approx 1.298$.}
\label{fig:u}
\end{figure}

Figure~\ref{fig:u} shows the variation of the LKDM parameter $u =
\frac{4\tilde{J}_{BB}S_B}{3\tilde{J}_{AB}S_A} =
\frac{4J_{BB}}{3J_{AB}}$ with the strength of the on-site Coulomb
interactions for the different lattice parameters and values of ${\cal
J}^H$ used in this work. The behavior of $u$ follows from the
corresponding trends in the coupling constants $J_{AB}$ and $J_{BB}$
discussed in the previous section.  Increasing $U_{A}$ decreases the
strength of $J_{AB}$ but leaves $J_{BB}$ more or less unchanged, and
thus increases the value of $u$. On the other hand both $J_{AB}$ and
$J_{BB}$ decrease with increasing $U_\text{Cr}$, but the decrease is
stronger for $J_{BB}$ and therefore $u$ decreases with increasing
$U_\text{Cr}$. Thus, the trends caused by the Hubbard parameters
corresponding to the two different magnetic sites are opposite to each
other.

As already pointed out in the previous section, changing the value of
the LSDA+$U$ parameter ${\cal J}^H$ and using different lattice
constants essentially just shifts the overall scale for the magnetic
coupling constants without altering their $U$ dependence. Therefore,
using the larger experimental volume decreases $u$ compared to the
value obtained at the theoretical lattice parameters due to the very
strong volume dependence of $J_{BB}$. Introducing the on-site Hund's
rule coupling ${\cal J}^H$ increases $u$, since $J_{BB}$ is stronger
affected by this and thus increases relative to $J_{AB}$.

For similar values of $U_A$ and $U_\text{Cr}$ the LKDM parameter $u$
is larger in MnCr$_2$O$_4$ than in CoCr$_2$O$_4$, except for ${\cal
J}^H = 1$\,eV at the theoretical lattice parameters, where they are
approximately equal. This is in contrast to what has been concluded by
fitting the experimental neutron spectra to the spiral spin structure
of the LKDM theory, which leads to the values $u$=1.6 for
MnCr$_2$O$_4$ and $u$=2.0 for
CoCr$_2$O$_4$,\cite{Hastings/Corliss:1962,Menyuk/Dwight/Wold:1964}
i.e. the fitted value for CoCr$_2$O$_4$ is significantly larger than the
value for MnCr$_2$O$_4$.

We now try to give a quantitative estimate of $u$ in the two systems.
The first question is whether using experimental or theoretical
lattice constants leads to more realistic magnetic coupling
constants. This question is not easy to answer in general. On the one
hand, the LSDA underestimation of the lattice constant can lead to an
overestimation of the magnetic coupling, since the cations are too
close together and can therefore interact stronger than at the
experimental volume. On the other hand, the indirect
cation-anion-cation interaction is intimately connected to the
chemical bonding.\cite{Goodenough:Book} If the larger experimental
lattice constant is used, this bonding is artificially suppressed and
the corresponding magnetic coupling is eventually underestimated. It
is therefore not obvious whether it is better to calculate the
coupling constants at the experimental or the theoretical lattice
parameters, but the two cases at least provide reasonable limits for
the magnetic coupling constants. We note that using the LSDA+$U$
method for the structural relaxation, usually leads to lattice
parameters that are in slightly better agreement with the experimental
values,\cite{Neaton_et_al:2005} which will decrease the corresponding
uncertainty in the magnetic coupling constants. In the present paper
we do not perform such structural relaxations for each combination of
the LSDA+$U$ parameters, in order to reduce the required computational
effort. In addition, this allows us to discuss the pure effect of $U$
and ${\cal J}^H$ on the magnetic coupling constants, without
contributions due to varying lattice parameters.

Fig.~\ref{fig:u} shows that for the physically reasonable parameters
$U_\text{Cr} = 3$\,eV, $U_A=$ 4-5\,eV, and ${\cal J}^H = 1$\,eV the
value of $u$ in CoCr$_2$O$_4$ calculated at the theoretical lattice
constant is slightly larger than the critical value $u'' \approx
1.298$, where within the LKDM theory the ferrimagnetic spiral
configuration becomes unstable. In MnCr$_2$O$_4$ the corresponding
value is about equal to $u''$. At the experimental lattice constants
the values of $u$ in both systems are smaller than at the theoretical
lattice constants, with the stronger effect in CoCr$_2$O$_4$, where
$u$ at the theoretical lattice constant is about equal to $u_0=8/9$,
the value below which, according to LKDM, a collinear ferrimagnetic
spin configuration is the ground state. In MnCr$_2$O$_4$ the value of
$u$ at the experimental lattice constant is between $u_0$ and
$u''$. Thus, in all cases the calculated values of $u$ are consistent
with the experimental evidence for noncollinear ordering.

Since in MnCr$_2$O$_4$ the calculation predicts a rather strong
$J_{AA}$, the validity of the LKDM theory is questionable for this
system, but for CoCr$_2$O$_4$, where the magnitude of $J_{AA}$ is
indeed significantly smaller than both $J_{AB}$ and $J_{BB}$, this
theory should at least be approximately correct. However, for
CoCr$_2$O$_4$ the calculated $u$ both at experimental and at the
theoretical lattice constant (and using physically reasonable values
for the LSDA+$U$ parameters) is still significantly smaller than the
value $u=2.0$ obtained by fitting the experimental data to the LKDM
theory.\cite{Menyuk/Dwight/Wold:1964} It would therefore be
interesting to study how the incorporation of $J_{AA}$ into a
generalized LKDM theory alters the conclusions drawn from the
experimental data. Obviously, a non-negligible $J_{AA}$ will further
destabilize the collinear N{\'e}el configuration, but the possible
influence of $J_{AA}$ on the ferrimagnetic spiral structure cannot be
obtained straightforwardly. Of course it cannot be fully excluded that
the discrepancy between the calculated value of $u$ for CoCr$_2$O$_4$
and the value extracted from the experimental data is caused by some
deficiencies of the LSDA+$U$ method. For example, it was shown in
Ref.~\onlinecite{Solovyev/Terakura:1998} that for MnO the LSDA+$U$
method does not offer enough degrees of freedom to correctly reproduce
both nearest and next nearest neighbor magnetic coupling constants.

Finally, we note that the fact that $J_{AA}$ is not negligible in
MnCr$_2$O$_4$ but has a significantly smaller magnitude than $J_{AB}$
and $J_{BB}$ in CoCr$_2$O$_4$ is compatible with the fact that the
overall agreement between the experimental data and the LKDM theory is
better for CoCr$_2$O$_4$ than for
MnCr$_2$O$_4$.\cite{Menyuk/Dwight/Wold:1964} However, a quantitative
discrepancy between the value of $u$ for CoCr$_2$O$_4$ calculated in
this work and the value derived from the experimental data remains.

\section{Summary and Conclusions}
\label{sec:summary}

In summary, we have presented a detailed LSDA+$U$ study of magnetic
coupling constants in the spinel systems CoCr$_2$O$_4$ and
MnCr$_2$O$_4$. We have found that the coupling between the $A$ site
cations, which is neglected in the classical theory of LKDM, is of
appreciable size in CoCr$_2$O$_4$ and definitely not negligible in
MnCr$_2$O$_4$. The calculated LKDM parameter $u$, which describes the
relative strength of the $BB$ coupling compared to the $AB$ coupling
and determines the nature of the ground state spin configuration in
the LKDM theory, is found to be smaller than the values obtained by
fitting experimental neutron data to the predictions of the LKDM
theory. It remains to be seen whether this discrepancy is caused by
the simplifications made in the LKDM theory, or whether it is due to
deficiencies of the LSDA+$U$ method used in our calculations.

In addition, we have shown that it is difficult, but possible, to
arrive at quantitative predictions of magnetic coupling constants
using the LSDA+$U$ method. In addition, by analyzing the $U$ and
${\cal J}^H$ dependence of the magnetic coupling constants it is
possible to identify the various interaction mechanisms contributing
to the overall magnetic coupling. The presence of two different
magnetic cations with different charge states and different anion
coordination, promotes the systems investigated in this work to a very
hard test case for the predictive capabilities of the LSDA+$U$
method. Nevertheless, some insight can be gained by a careful analysis
of all methodological uncertainties, and the magnitudes of the
magnetic coupling constants can be determined to a degree of accuracy
that allows to establish important trends and predict the correct
order of magnitude for the corresponding effects.

\begin{acknowledgments}
  
  C.E. thanks Craig Fennie, Ram Seshadri, Nicola Spaldin, and Andrew
  Millis for useful discussions.  This work was supported by the NSF's
  \emph{Chemical Bonding Centers} program, Grant No. CHE-0434567 and
  by the MRSEC Program of the NSF under the award number
  DMR-0213574. We also made use of central facilities provided by
  NSF-MRSEC Award No. DMR-0520415.

\end{acknowledgments}

\bibliography{references.bib}

\begin{thebibliography}{50}
\expandafter\ifx\csname natexlab\endcsname\relax\def\natexlab#1{#1}\fi
\expandafter\ifx\csname bibnamefont\endcsname\relax
  \def\bibnamefont#1{#1}\fi
\expandafter\ifx\csname bibfnamefont\endcsname\relax
  \def\bibfnamefont#1{#1}\fi
\expandafter\ifx\csname citenamefont\endcsname\relax
  \def\citenamefont#1{#1}\fi
\expandafter\ifx\csname url\endcsname\relax
  \def\url#1{\texttt{#1}}\fi
\expandafter\ifx\csname urlprefix\endcsname\relax\def\urlprefix{URL }\fi
\providecommand{\bibinfo}[2]{#2}
\providecommand{\eprint}[2][]{\url{#2}}

\bibitem[{\citenamefont{Ramirez}(2001)}]{Ramirez:2001}
\bibinfo{author}{\bibfnamefont{A.~P.} \bibnamefont{Ramirez}}, in
  \emph{\bibinfo{booktitle}{Handbook of magnetic materials}}, edited by
  \bibinfo{editor}{\bibfnamefont{K.~H.~J.} \bibnamefont{Bushow}}
  (\bibinfo{publisher}{Elsevier Science B.~V.}, \bibinfo{year}{2001}),
  vol.~\bibinfo{volume}{13}, chap.~\bibinfo{chapter}{4}.

\bibitem[{\citenamefont{Ramirez}(2003)}]{Ramirez:2003}
\bibinfo{author}{\bibfnamefont{A.~P.} \bibnamefont{Ramirez}},
  \bibinfo{journal}{Nature (London)} \textbf{\bibinfo{volume}{421}},
  \bibinfo{pages}{483} (\bibinfo{year}{2003}).

\bibitem[{\citenamefont{Lee et~al.}(2000)\citenamefont{Lee, Broholm, Kim,
  {Ratcliff II}, and Cheong}}]{Lee_et_al:2000}
\bibinfo{author}{\bibfnamefont{S.-H.} \bibnamefont{Lee}},
  \bibinfo{author}{\bibfnamefont{C.}~\bibnamefont{Broholm}},
  \bibinfo{author}{\bibfnamefont{T.~H.} \bibnamefont{Kim}},
  \bibinfo{author}{\bibfnamefont{W.}~\bibnamefont{{Ratcliff II}}},
  \bibnamefont{and} \bibinfo{author}{\bibfnamefont{S.-W.}
  \bibnamefont{Cheong}}, \bibinfo{journal}{Phys. Rev. Lett.}
  \textbf{\bibinfo{volume}{84}}, \bibinfo{pages}{3718} (\bibinfo{year}{2000}).

\bibitem[{\citenamefont{Chung et~al.}(2005)\citenamefont{Chung, Matsuda, Lee,
  Kakurai, Ueda, Sato, Takagi, Hong, and Park}}]{Chung_et_al:2005}
\bibinfo{author}{\bibfnamefont{J.-H.} \bibnamefont{Chung}},
  \bibinfo{author}{\bibfnamefont{M.}~\bibnamefont{Matsuda}},
  \bibinfo{author}{\bibfnamefont{S.-H.} \bibnamefont{Lee}},
  \bibinfo{author}{\bibfnamefont{K.}~\bibnamefont{Kakurai}},
  \bibinfo{author}{\bibfnamefont{H.}~\bibnamefont{Ueda}},
  \bibinfo{author}{\bibfnamefont{T.~J.} \bibnamefont{Sato}},
  \bibinfo{author}{\bibfnamefont{H.}~\bibnamefont{Takagi}},
  \bibinfo{author}{\bibfnamefont{K.-P.} \bibnamefont{Hong}}, \bibnamefont{and}
  \bibinfo{author}{\bibfnamefont{S.}~\bibnamefont{Park}},
  \bibinfo{journal}{Phys. Rev. Lett.} \textbf{\bibinfo{volume}{95}},
  \bibinfo{pages}{247204} (\bibinfo{year}{2005}).

\bibitem[{\citenamefont{Sushkov et~al.}(2005)\citenamefont{Sushkov,
  Tchernyshyov, {Ratcliff II}, Cheong, and Drew}}]{Sushkov_et_al:2005}
\bibinfo{author}{\bibfnamefont{A.~B.} \bibnamefont{Sushkov}},
  \bibinfo{author}{\bibfnamefont{O.}~\bibnamefont{Tchernyshyov}},
  \bibinfo{author}{\bibfnamefont{W.}~\bibnamefont{{Ratcliff II}}},
  \bibinfo{author}{\bibfnamefont{S.~W.} \bibnamefont{Cheong}},
  \bibnamefont{and} \bibinfo{author}{\bibfnamefont{H.~D.} \bibnamefont{Drew}},
  \bibinfo{journal}{Phys. Rev. Lett.} \textbf{\bibinfo{volume}{94}},
  \bibinfo{pages}{137202} (\bibinfo{year}{2005}).

\bibitem[{\citenamefont{Fennie and Rabe}(2006)}]{Fennie/Rabe:2006}
\bibinfo{author}{\bibfnamefont{C.~J.} \bibnamefont{Fennie}} \bibnamefont{and}
  \bibinfo{author}{\bibfnamefont{K.~M.} \bibnamefont{Rabe}},
  \bibinfo{journal}{Phys. Rev. Lett.} \textbf{\bibinfo{volume}{96}},
  \bibinfo{pages}{205505} (\bibinfo{year}{2006}).

\bibitem[{\citenamefont{Hastings and Corliss}(1962)}]{Hastings/Corliss:1962}
\bibinfo{author}{\bibfnamefont{J.~M.} \bibnamefont{Hastings}} \bibnamefont{and}
  \bibinfo{author}{\bibfnamefont{L.~M.} \bibnamefont{Corliss}},
  \bibinfo{journal}{Phys. Rev.} \textbf{\bibinfo{volume}{126}},
  \bibinfo{pages}{556} (\bibinfo{year}{1962}).

\bibitem[{\citenamefont{Menyuk et~al.}(1964)\citenamefont{Menyuk, Dwight, and
  Wold}}]{Menyuk/Dwight/Wold:1964}
\bibinfo{author}{\bibfnamefont{N.}~\bibnamefont{Menyuk}},
  \bibinfo{author}{\bibfnamefont{K.}~\bibnamefont{Dwight}}, \bibnamefont{and}
  \bibinfo{author}{\bibfnamefont{A.}~\bibnamefont{Wold}},
  \bibinfo{journal}{J.~Phys. (Paris)} \textbf{\bibinfo{volume}{25}},
  \bibinfo{pages}{528} (\bibinfo{year}{1964}).

\bibitem[{\citenamefont{Kimura et~al.}(2003)\citenamefont{Kimura, Goto,
  Shintani, Ishizaka, Arima, and Tokura}}]{Kimura_et_al_Nature:2003}
\bibinfo{author}{\bibfnamefont{T.}~\bibnamefont{Kimura}},
  \bibinfo{author}{\bibfnamefont{T.}~\bibnamefont{Goto}},
  \bibinfo{author}{\bibfnamefont{H.}~\bibnamefont{Shintani}},
  \bibinfo{author}{\bibfnamefont{K.}~\bibnamefont{Ishizaka}},
  \bibinfo{author}{\bibfnamefont{T.}~\bibnamefont{Arima}}, \bibnamefont{and}
  \bibinfo{author}{\bibfnamefont{Y.}~\bibnamefont{Tokura}},
  \bibinfo{journal}{Nature (London)} \textbf{\bibinfo{volume}{426}},
  \bibinfo{pages}{55} (\bibinfo{year}{2003}).

\bibitem[{\citenamefont{Lawes et~al.}(2005)\citenamefont{Lawes, Harris, Kimura,
  Rogado, Cava, Aharony, Entin-Wohlmann, Yildirim, Kenzelmann, Broholm
  et~al.}}]{Lawes_et_al:2005}
\bibinfo{author}{\bibfnamefont{G.}~\bibnamefont{Lawes}},
  \bibinfo{author}{\bibfnamefont{A.~B.} \bibnamefont{Harris}},
  \bibinfo{author}{\bibfnamefont{T.}~\bibnamefont{Kimura}},
  \bibinfo{author}{\bibfnamefont{N.}~\bibnamefont{Rogado}},
  \bibinfo{author}{\bibfnamefont{R.~J.} \bibnamefont{Cava}},
  \bibinfo{author}{\bibfnamefont{A.}~\bibnamefont{Aharony}},
  \bibinfo{author}{\bibfnamefont{O.}~\bibnamefont{Entin-Wohlmann}},
  \bibinfo{author}{\bibfnamefont{T.}~\bibnamefont{Yildirim}},
  \bibinfo{author}{\bibfnamefont{M.}~\bibnamefont{Kenzelmann}},
  \bibinfo{author}{\bibfnamefont{C.}~\bibnamefont{Broholm}},
  \bibnamefont{et~al.}, \bibinfo{journal}{Phys. Rev. Lett.}
  \textbf{\bibinfo{volume}{95}}, \bibinfo{pages}{087205}
  (\bibinfo{year}{2005}).

\bibitem[{\citenamefont{Lawes et~al.}(2006)\citenamefont{Lawes, Melot, Page,
  Ederer, Hayward, Proffen, and Seshadri}}]{Lawes_et_al:2006}
\bibinfo{author}{\bibfnamefont{G.}~\bibnamefont{Lawes}},
  \bibinfo{author}{\bibfnamefont{B.}~\bibnamefont{Melot}},
  \bibinfo{author}{\bibfnamefont{K.}~\bibnamefont{Page}},
  \bibinfo{author}{\bibfnamefont{C.}~\bibnamefont{Ederer}},
  \bibinfo{author}{\bibfnamefont{M.~A.} \bibnamefont{Hayward}},
  \bibinfo{author}{\bibfnamefont{T.}~\bibnamefont{Proffen}}, \bibnamefont{and}
  \bibinfo{author}{\bibfnamefont{R.}~\bibnamefont{Seshadri}},
  \bibinfo{journal}{Phys. Rev. B} \textbf{\bibinfo{volume}{74}},
  \bibinfo{pages}{024413} (\bibinfo{year}{2006}).

\bibitem[{\citenamefont{Yamasaki et~al.}(2006)\citenamefont{Yamasaki, Miyasaka,
  Kaneko, He, Arima, and Tokura}}]{Yamasaki_et_al:2006}
\bibinfo{author}{\bibfnamefont{Y.}~\bibnamefont{Yamasaki}},
  \bibinfo{author}{\bibfnamefont{S.}~\bibnamefont{Miyasaka}},
  \bibinfo{author}{\bibfnamefont{Y.}~\bibnamefont{Kaneko}},
  \bibinfo{author}{\bibfnamefont{J.-P.} \bibnamefont{He}},
  \bibinfo{author}{\bibfnamefont{T.}~\bibnamefont{Arima}}, \bibnamefont{and}
  \bibinfo{author}{\bibfnamefont{Y.}~\bibnamefont{Tokura}},
  \bibinfo{journal}{Phys. Rev. Lett.} \textbf{\bibinfo{volume}{96}},
  \bibinfo{pages}{207204} (\bibinfo{year}{2006}).

\bibitem[{\citenamefont{Lyons et~al.}(1962)\citenamefont{Lyons, Kaplan, Dwight,
  and Menyuk}}]{Lyons_et_al:1962}
\bibinfo{author}{\bibfnamefont{D.~H.} \bibnamefont{Lyons}},
  \bibinfo{author}{\bibfnamefont{T.~A.} \bibnamefont{Kaplan}},
  \bibinfo{author}{\bibfnamefont{K.}~\bibnamefont{Dwight}}, \bibnamefont{and}
  \bibinfo{author}{\bibfnamefont{N.}~\bibnamefont{Menyuk}},
  \bibinfo{journal}{Phys. Rev.} \textbf{\bibinfo{volume}{126}},
  \bibinfo{pages}{540} (\bibinfo{year}{1962}).

\bibitem[{foo()}]{footnote1}
\bibinfo{note}{In this nomenclature the Heisenberg interaction energy is
  expressed as $E_{ij} = - 2 \tilde{J}_{ij} \vec{S}_i \cdot \vec{S}_j$, where
  $\vec{S}_i$ is a classical vector of length $S_i$, the total spin of ion
  $i$.}

\bibitem[{\citenamefont{Tomiyasu et~al.}(2004)\citenamefont{Tomiyasu, Fukunaga,
  and Suzuki}}]{Tomiyasu/Fukunaga/Suzuki:2004}
\bibinfo{author}{\bibfnamefont{K.}~\bibnamefont{Tomiyasu}},
  \bibinfo{author}{\bibfnamefont{J.}~\bibnamefont{Fukunaga}}, \bibnamefont{and}
  \bibinfo{author}{\bibfnamefont{H.}~\bibnamefont{Suzuki}},
  \bibinfo{journal}{Phys. Rev. B} \textbf{\bibinfo{volume}{70}},
  \bibinfo{pages}{214434} (\bibinfo{year}{2004}).

\bibitem[{\citenamefont{Jones and Gunnarsson}(1989)}]{Jones/Gunnarsson:1989}
\bibinfo{author}{\bibfnamefont{R.~O.} \bibnamefont{Jones}} \bibnamefont{and}
  \bibinfo{author}{\bibfnamefont{O.}~\bibnamefont{Gunnarsson}},
  \bibinfo{journal}{Rev. Mod. Phys.} \textbf{\bibinfo{volume}{61}},
  \bibinfo{pages}{689} (\bibinfo{year}{1989}).

\bibitem[{\citenamefont{Anisimov
  et~al.}(1997{\natexlab{a}})\citenamefont{Anisimov, Aryasetiawan, and
  Liechtenstein}}]{Anisimov/Aryatesiawan/Liechtenstein:1997}
\bibinfo{author}{\bibfnamefont{V.~I.} \bibnamefont{Anisimov}},
  \bibinfo{author}{\bibfnamefont{F.}~\bibnamefont{Aryasetiawan}},
  \bibnamefont{and} \bibinfo{author}{\bibfnamefont{A.~I.}
  \bibnamefont{Liechtenstein}}, \bibinfo{journal}{J.~Phys.: Condens. Matter}
  \textbf{\bibinfo{volume}{9}}, \bibinfo{pages}{767}
  (\bibinfo{year}{1997}{\natexlab{a}}).

\bibitem[{\citenamefont{Solovyev and Terakura}(1998)}]{Solovyev/Terakura:1998}
\bibinfo{author}{\bibfnamefont{I.~V.} \bibnamefont{Solovyev}} \bibnamefont{and}
  \bibinfo{author}{\bibfnamefont{K.}~\bibnamefont{Terakura}},
  \bibinfo{journal}{Phys. Rev. B} \textbf{\bibinfo{volume}{58}},
  \bibinfo{pages}{15496} (\bibinfo{year}{1998}).

\bibitem[{\citenamefont{Yaresko et~al.}(2000)\citenamefont{Yaresko, Antonov,
  Eschrig, Thalmaier, and Fulde}}]{Yaresko_et_al:2000}
\bibinfo{author}{\bibfnamefont{A.~N.} \bibnamefont{Yaresko}},
  \bibinfo{author}{\bibfnamefont{V.~N.} \bibnamefont{Antonov}},
  \bibinfo{author}{\bibfnamefont{H.}~\bibnamefont{Eschrig}},
  \bibinfo{author}{\bibfnamefont{P.}~\bibnamefont{Thalmaier}},
  \bibnamefont{and} \bibinfo{author}{\bibfnamefont{P.}~\bibnamefont{Fulde}},
  \bibinfo{journal}{Phys. Rev. B} \textbf{\bibinfo{volume}{62}},
  \bibinfo{pages}{15538} (\bibinfo{year}{2000}).

\bibitem[{\citenamefont{Baettig et~al.}(2005)\citenamefont{Baettig, Ederer, and
  Spaldin}}]{Baettig/Ederer/Spaldin:2005}
\bibinfo{author}{\bibfnamefont{P.}~\bibnamefont{Baettig}},
  \bibinfo{author}{\bibfnamefont{C.}~\bibnamefont{Ederer}}, \bibnamefont{and}
  \bibinfo{author}{\bibfnamefont{N.~A.} \bibnamefont{Spaldin}},
  \bibinfo{journal}{Phys. Rev. B} \textbf{\bibinfo{volume}{72}},
  \bibinfo{pages}{214105} (\bibinfo{year}{2005}).

\bibitem[{\citenamefont{Nov{\'a}k and Rusz}(2005)}]{Novak/Rusz:2005}
\bibinfo{author}{\bibfnamefont{P.}~\bibnamefont{Nov{\'a}k}} \bibnamefont{and}
  \bibinfo{author}{\bibfnamefont{J.}~\bibnamefont{Rusz}},
  \bibinfo{journal}{Phys. Rev. B} \textbf{\bibinfo{volume}{71}},
  \bibinfo{pages}{184433} (\bibinfo{year}{2005}).

\bibitem[{\citenamefont{Anisimov et~al.}(1991)\citenamefont{Anisimov, Zaanen,
  and Andersen}}]{Anisimov/Zaanen/Andersen:1991}
\bibinfo{author}{\bibfnamefont{V.~I.} \bibnamefont{Anisimov}},
  \bibinfo{author}{\bibfnamefont{J.}~\bibnamefont{Zaanen}}, \bibnamefont{and}
  \bibinfo{author}{\bibfnamefont{O.~K.} \bibnamefont{Andersen}},
  \bibinfo{journal}{Phys. Rev. B} \textbf{\bibinfo{volume}{44}},
  \bibinfo{pages}{943} (\bibinfo{year}{1991}).

\bibitem[{\citenamefont{Anisimov
  et~al.}(1997{\natexlab{b}})\citenamefont{Anisimov, Potaryaev, Korotin,
  Anokhin, and Kotliar}}]{Anisimov_et_al:1997}
\bibinfo{author}{\bibfnamefont{V.~I.} \bibnamefont{Anisimov}},
  \bibinfo{author}{\bibfnamefont{A.~I.} \bibnamefont{Potaryaev}},
  \bibinfo{author}{\bibfnamefont{M.~A.} \bibnamefont{Korotin}},
  \bibinfo{author}{\bibfnamefont{A.~O.} \bibnamefont{Anokhin}},
  \bibnamefont{and} \bibinfo{author}{\bibfnamefont{G.}~\bibnamefont{Kotliar}},
  \bibinfo{journal}{J.~Phys.: Condens. Matter} \textbf{\bibinfo{volume}{9}},
  \bibinfo{pages}{7359} (\bibinfo{year}{1997}{\natexlab{b}}).

\bibitem[{\citenamefont{Solovyev et~al.}(1994)\citenamefont{Solovyev,
  Dederichs, and Anisimov}}]{Solovyev/Dederichs/Anisimov:1994}
\bibinfo{author}{\bibfnamefont{I.~V.} \bibnamefont{Solovyev}},
  \bibinfo{author}{\bibfnamefont{P.~H.} \bibnamefont{Dederichs}},
  \bibnamefont{and} \bibinfo{author}{\bibfnamefont{V.~I.}
  \bibnamefont{Anisimov}}, \bibinfo{journal}{Phys. Rev. B}
  \textbf{\bibinfo{volume}{50}}, \bibinfo{pages}{16861} (\bibinfo{year}{1994}).

\bibitem[{\citenamefont{Czy{\.z}yk and Sawatzky}(1994)}]{Czyzyk/Sawatzky:1994}
\bibinfo{author}{\bibfnamefont{M.~T.} \bibnamefont{Czy{\.z}yk}}
  \bibnamefont{and} \bibinfo{author}{\bibfnamefont{G.~A.}
  \bibnamefont{Sawatzky}}, \bibinfo{journal}{Phys. Rev. B}
  \textbf{\bibinfo{volume}{49}}, \bibinfo{pages}{14211} (\bibinfo{year}{1994}).

\bibitem[{\citenamefont{Dederichs et~al.}(1984)\citenamefont{Dederichs,
  Bl{\"u}gel, Zeller, and Akai}}]{Dederichs_et_al:1984}
\bibinfo{author}{\bibfnamefont{P.~H.} \bibnamefont{Dederichs}},
  \bibinfo{author}{\bibfnamefont{S.}~\bibnamefont{Bl{\"u}gel}},
  \bibinfo{author}{\bibfnamefont{R.}~\bibnamefont{Zeller}}, \bibnamefont{and}
  \bibinfo{author}{\bibfnamefont{H.}~\bibnamefont{Akai}},
  \bibinfo{journal}{Phys. Rev. Lett.} \textbf{\bibinfo{volume}{53}},
  \bibinfo{pages}{2512} (\bibinfo{year}{1984}).

\bibitem[{\citenamefont{Hybertsen et~al.}(1989)\citenamefont{Hybertsen,
  Schl{\"u}ter, and Christensen}}]{Hybertsen/Schlueter/Christensen:1989}
\bibinfo{author}{\bibfnamefont{M.~S.} \bibnamefont{Hybertsen}},
  \bibinfo{author}{\bibfnamefont{M.}~\bibnamefont{Schl{\"u}ter}},
  \bibnamefont{and} \bibinfo{author}{\bibfnamefont{N.~E.}
  \bibnamefont{Christensen}}, \bibinfo{journal}{Phys. Rev. B}
  \textbf{\bibinfo{volume}{39}}, \bibinfo{pages}{9028} (\bibinfo{year}{1989}).

\bibitem[{\citenamefont{Solovyev et~al.}(1996)\citenamefont{Solovyev, Hamada,
  and Terakura}}]{Solovyev/Hamada/Terakura:1996}
\bibinfo{author}{\bibfnamefont{I.}~\bibnamefont{Solovyev}},
  \bibinfo{author}{\bibfnamefont{N.}~\bibnamefont{Hamada}}, \bibnamefont{and}
  \bibinfo{author}{\bibfnamefont{K.}~\bibnamefont{Terakura}},
  \bibinfo{journal}{Phys. Rev. B} \textbf{\bibinfo{volume}{53}},
  \bibinfo{pages}{7158} (\bibinfo{year}{1996}).

\bibitem[{\citenamefont{Pickett et~al.}(1998)\citenamefont{Pickett, Erwin, and
  Ethridge}}]{Pickett/Erwin/Ethridge:1998}
\bibinfo{author}{\bibfnamefont{W.~E.} \bibnamefont{Pickett}},
  \bibinfo{author}{\bibfnamefont{S.~C.} \bibnamefont{Erwin}}, \bibnamefont{and}
  \bibinfo{author}{\bibfnamefont{E.~C.} \bibnamefont{Ethridge}},
  \bibinfo{journal}{Phys. Rev. B} \textbf{\bibinfo{volume}{58}},
  \bibinfo{pages}{1201} (\bibinfo{year}{1998}).

\bibitem[{\citenamefont{Cococcioni and
  de~Gironcoli}(2005)}]{Cococcioni/Gironcoli:2005}
\bibinfo{author}{\bibfnamefont{M.}~\bibnamefont{Cococcioni}} \bibnamefont{and}
  \bibinfo{author}{\bibfnamefont{S.}~\bibnamefont{de~Gironcoli}},
  \bibinfo{journal}{Phys. Rev. B} \textbf{\bibinfo{volume}{71}},
  \bibinfo{pages}{035105} (\bibinfo{year}{2005}).

\bibitem[{\citenamefont{Anisimov and
  Gunnarsson}(1991)}]{Anisimov/Gunnarsson:1991}
\bibinfo{author}{\bibfnamefont{V.~I.} \bibnamefont{Anisimov}} \bibnamefont{and}
  \bibinfo{author}{\bibfnamefont{O.}~\bibnamefont{Gunnarsson}},
  \bibinfo{journal}{Phys. Rev. B} \textbf{\bibinfo{volume}{43}},
  \bibinfo{pages}{7570} (\bibinfo{year}{1991}).

\bibitem[{\citenamefont{Fennie and Rabe}(2005)}]{Fennie/Rabe_CCS:2005}
\bibinfo{author}{\bibfnamefont{C.~J.} \bibnamefont{Fennie}} \bibnamefont{and}
  \bibinfo{author}{\bibfnamefont{K.~M.} \bibnamefont{Rabe}},
  \bibinfo{journal}{Phys. Rev. B} \textbf{\bibinfo{volume}{72}},
  \bibinfo{pages}{214123} (\bibinfo{year}{2005}).

\bibitem[{\citenamefont{Sawada et~al.}(1997)\citenamefont{Sawada, Morikawa,
  Terakura, and Hamada}}]{Sawada_et_al:1997}
\bibinfo{author}{\bibfnamefont{H.}~\bibnamefont{Sawada}},
  \bibinfo{author}{\bibfnamefont{Y.}~\bibnamefont{Morikawa}},
  \bibinfo{author}{\bibfnamefont{K.}~\bibnamefont{Terakura}}, \bibnamefont{and}
  \bibinfo{author}{\bibfnamefont{N.}~\bibnamefont{Hamada}},
  \bibinfo{journal}{Phys. Rev. B} \textbf{\bibinfo{volume}{56}},
  \bibinfo{pages}{12154} (\bibinfo{year}{1997}).

\bibitem[{\citenamefont{Dudarev et~al.}(1998)\citenamefont{Dudarev, Botton,
  Savrasov, Humphreys, and Sutton}}]{Dudarev_et_al:1998}
\bibinfo{author}{\bibfnamefont{S.~L.} \bibnamefont{Dudarev}},
  \bibinfo{author}{\bibfnamefont{G.~A.} \bibnamefont{Botton}},
  \bibinfo{author}{\bibfnamefont{S.~Y.} \bibnamefont{Savrasov}},
  \bibinfo{author}{\bibfnamefont{C.~J.} \bibnamefont{Humphreys}},
  \bibnamefont{and} \bibinfo{author}{\bibfnamefont{A.~P.}
  \bibnamefont{Sutton}}, \bibinfo{journal}{Phys. Rev. B}
  \textbf{\bibinfo{volume}{57}}, \bibinfo{pages}{1505} (\bibinfo{year}{1998}).

\bibitem[{\citenamefont{Ederer and Spaldin}(2005)}]{Ederer/Spaldin_2:2005}
\bibinfo{author}{\bibfnamefont{C.}~\bibnamefont{Ederer}} \bibnamefont{and}
  \bibinfo{author}{\bibfnamefont{N.~A.} \bibnamefont{Spaldin}},
  \bibinfo{journal}{Phys. Rev. B} \textbf{\bibinfo{volume}{71}},
  \bibinfo{pages}{224103} (\bibinfo{year}{2005}).

\bibitem[{\citenamefont{Liechtenstein et~al.}(1984)\citenamefont{Liechtenstein,
  Katsnelson, and Gubanov}}]{Liechtenstein/Katsnelson/Gubanov:1984}
\bibinfo{author}{\bibfnamefont{A.~I.} \bibnamefont{Liechtenstein}},
  \bibinfo{author}{\bibfnamefont{M.~I.} \bibnamefont{Katsnelson}},
  \bibnamefont{and} \bibinfo{author}{\bibfnamefont{V.~A.}
  \bibnamefont{Gubanov}}, \bibinfo{journal}{J.~Phys. F: Met. Phys.}
  \textbf{\bibinfo{volume}{14}}, \bibinfo{pages}{L125} (\bibinfo{year}{1984}).

\bibitem[{\citenamefont{Bl{\"o}chl}(1994)}]{Bloechl:1994}
\bibinfo{author}{\bibfnamefont{P.~E.} \bibnamefont{Bl{\"o}chl}},
  \bibinfo{journal}{Phys. Rev. B} \textbf{\bibinfo{volume}{50}},
  \bibinfo{pages}{17953} (\bibinfo{year}{1994}).

\bibitem[{\citenamefont{Kresse and
  Furthm{\"u}ller}(1996)}]{Kresse/Furthmueller_PRB:1996}
\bibinfo{author}{\bibfnamefont{G.}~\bibnamefont{Kresse}} \bibnamefont{and}
  \bibinfo{author}{\bibfnamefont{J.}~\bibnamefont{Furthm{\"u}ller}},
  \bibinfo{journal}{Phys. Rev. B} \textbf{\bibinfo{volume}{54}},
  \bibinfo{pages}{11169} (\bibinfo{year}{1996}).

\bibitem[{\citenamefont{Kresse and Joubert}(1999)}]{Kresse/Joubert:1999}
\bibinfo{author}{\bibfnamefont{G.}~\bibnamefont{Kresse}} \bibnamefont{and}
  \bibinfo{author}{\bibfnamefont{D.}~\bibnamefont{Joubert}},
  \bibinfo{journal}{Phys. Rev. B} \textbf{\bibinfo{volume}{59}},
  \bibinfo{pages}{1758} (\bibinfo{year}{1999}).

\bibitem[{\citenamefont{Wimmer et~al.}(1981)\citenamefont{Wimmer, Krakauer,
  Weinert, and Freeman}}]{Wimmer:1981}
\bibinfo{author}{\bibfnamefont{E.}~\bibnamefont{Wimmer}},
  \bibinfo{author}{\bibfnamefont{H.}~\bibnamefont{Krakauer}},
  \bibinfo{author}{\bibfnamefont{M.}~\bibnamefont{Weinert}}, \bibnamefont{and}
  \bibinfo{author}{\bibfnamefont{A.~J.} \bibnamefont{Freeman}},
  \bibinfo{journal}{Phys. Rev. B} \textbf{\bibinfo{volume}{24}},
  \bibinfo{pages}{864} (\bibinfo{year}{1981}).

\bibitem[{\citenamefont{Blaha et~al.}(1990)\citenamefont{Blaha, Schwarz,
  Sorantin, and Trickey}}]{Blaha:1990}
\bibinfo{author}{\bibfnamefont{P.}~\bibnamefont{Blaha}},
  \bibinfo{author}{\bibfnamefont{K.}~\bibnamefont{Schwarz}},
  \bibinfo{author}{\bibfnamefont{P.}~\bibnamefont{Sorantin}}, \bibnamefont{and}
  \bibinfo{author}{\bibfnamefont{S.~B.} \bibnamefont{Trickey}},
  \bibinfo{journal}{Comput. Phys. Commun.} \textbf{\bibinfo{volume}{59}},
  \bibinfo{pages}{399} (\bibinfo{year}{1990}).

\bibitem[{Ram()}]{Ram_private}
\bibinfo{note}{R.~Seshadri, private communication}.

\bibitem[{\citenamefont{Neaton et~al.}(2005)\citenamefont{Neaton, Ederer,
  Waghmare, Spaldin, and Rabe}}]{Neaton_et_al:2005}
\bibinfo{author}{\bibfnamefont{J.~B.} \bibnamefont{Neaton}},
  \bibinfo{author}{\bibfnamefont{C.}~\bibnamefont{Ederer}},
  \bibinfo{author}{\bibfnamefont{U.~V.} \bibnamefont{Waghmare}},
  \bibinfo{author}{\bibfnamefont{N.~A.} \bibnamefont{Spaldin}},
  \bibnamefont{and} \bibinfo{author}{\bibfnamefont{K.~M.} \bibnamefont{Rabe}},
  \bibinfo{journal}{Phys. Rev. B} \textbf{\bibinfo{volume}{71}},
  \bibinfo{pages}{014113} (\bibinfo{year}{2005}).

\bibitem[{\citenamefont{Goodenough}(1963)}]{Goodenough:Book}
\bibinfo{author}{\bibfnamefont{J.~B.} \bibnamefont{Goodenough}},
  \emph{\bibinfo{title}{Magnetism and the Chemical Bond}}
  (\bibinfo{publisher}{Interscience Publishers, New York},
  \bibinfo{year}{1963}).

\bibitem[{\citenamefont{Anderson}(1963)}]{Anderson:1963}
\bibinfo{author}{\bibfnamefont{P.~W.} \bibnamefont{Anderson}}, in
  \emph{\bibinfo{booktitle}{Magnetism}}, edited by
  \bibinfo{editor}{\bibfnamefont{G.~T.} \bibnamefont{Rado}} \bibnamefont{and}
  \bibinfo{editor}{\bibfnamefont{H.}~\bibnamefont{Suhl}}
  (\bibinfo{publisher}{Academic Press}, \bibinfo{year}{1963}),
  vol.~\bibinfo{volume}{1}, chap.~\bibinfo{chapter}{2}, pp.
  \bibinfo{pages}{25--83}.

\bibitem[{\citenamefont{Wickham and
  Goodenough}(1959)}]{Wickham/Goodenough:1959}
\bibinfo{author}{\bibfnamefont{D.~G.} \bibnamefont{Wickham}} \bibnamefont{and}
  \bibinfo{author}{\bibfnamefont{J.~B.} \bibnamefont{Goodenough}},
  \bibinfo{journal}{Phys. Rev.} \textbf{\bibinfo{volume}{115}},
  \bibinfo{pages}{1156} (\bibinfo{year}{1959}).

\bibitem[{\citenamefont{Singh et~al.}(2002)\citenamefont{Singh, Gupta, and
  Gupta}}]{Singh/Gupta/Gupta:2002}
\bibinfo{author}{\bibfnamefont{D.~J.} \bibnamefont{Singh}},
  \bibinfo{author}{\bibfnamefont{M.}~\bibnamefont{Gupta}}, \bibnamefont{and}
  \bibinfo{author}{\bibfnamefont{R.}~\bibnamefont{Gupta}},
  \bibinfo{journal}{Phys. Rev. B} \textbf{\bibinfo{volume}{65}},
  \bibinfo{pages}{064432} (\bibinfo{year}{2002}).

\bibitem[{\citenamefont{Dwight and Menyuk}(1967)}]{Dwight/Menyuk:1967}
\bibinfo{author}{\bibfnamefont{K.}~\bibnamefont{Dwight}} \bibnamefont{and}
  \bibinfo{author}{\bibfnamefont{N.}~\bibnamefont{Menyuk}},
  \bibinfo{journal}{Phys. Rev.} \textbf{\bibinfo{volume}{163}},
  \bibinfo{pages}{435} (\bibinfo{year}{1967}).

\bibitem[{dis()}]{distances}
\bibinfo{note}{All values refer to CoCr$_2$O$_4$ and experimental lattice
  parameters.}

\bibitem[{\citenamefont{Fennie}()}]{Craig:unpublished}
\bibinfo{author}{\bibfnamefont{C.~J.} \bibnamefont{Fennie}}, \bibinfo{note}{in
  preparation}.

\end{thebibliography}

\end{document}